\documentstyle[prb,aps,amssymb,preprint]{revtex}

\begin{document}
\draft

\date{Received \today}

\title{
Quantum Fluctuations in the spiral phase of the Hubbard Model}
\author{ C. Zhou$^{a,b}$ and H. J. Schulz$^a$}
\address{$^a$Laboratoire de Physique des Solides, Universit\'e Paris--Sud,
91405 Orsay, France}
\address{$^b$Laboratoire de Physique Quantique, Universit\'e Paul Sabatier,
31062 Toulouse, France}

\maketitle

\begin{abstract}
We study the magnetic excitations in the spiral phase of the
two--dimensional Hubbard model using a functional integral method. Spin
waves are strongly renormalized and a line of near--zeros is observed in the
spectrum around the spiral pitch $\pm{\bf Q}$.  The possibility of
disordered spiral states is examined by studying the one--loop corrections
to the spiral order parameter.  We also show that the spiral phase presents
an intrinsic instability towards an inhomogeneous state (phase separation,
CDW, ...) at weak doping.  Though phase separation is suppressed by weak
long--range
Coulomb
interactions, the CDW instability only disappears for sufficiently
strong Coulomb interaction.
\end{abstract}


\narrowtext

The proximity of superconductivity and antiferromagnetism observed in the
copper--oxide superconductors has led to intensive studies on the magnetic
properties and their interplay with charge dynamics in the CuO$_2$
planes. The insulating (half--filled) compounds are N\'eel
antiferromagnets, and upon doping strong antiferromagnetic fluctuations
persist both in the normal and superconducting phases of the copper
oxides. In particular, recent neutron data on
La$_{\text{2-x}}$Sr$_{\text{x}}$CuO$_{\text{4}}$ have revealed
incommensurate fluctuations peaks in the dynamical susceptibility
$\chi^{\prime\prime}({\bf q},
\omega).$\cite{cheong_neutron_incom,mason_neutron_incom,mason_neutron_incom1}
The effect of finite doping in the antiferromagnetic background has thus
been extensively studied within the framework of the Hubbard model or its
strong--coupling derivations.

Within the Hartree--Fock approximation, doping induces an incommensurate
magnetic order since the magnetic susceptibility then peaks at
incommensurate wave vectors instead of ${\bf Q}_0=(\pi,\pi)$ as in the half
filling.\cite{schulz_incomm,tremblay_neutron}  In particular, it has
become
clear that doping initially gives rise to an ({\it inhomogeneous})
antiferromagnetic insulator before a conducting phase is eventually reached
upon further doping.\cite{schulz_incomm,schulz_hf_incomm}  For strong
correlations, mean field theories based on the t--J model propose a spiral
magnetic
order.\cite{shraiman_tj_spiral1,shraiman_tj_spiral2,kane_tj_spiral,sarker_spiral1}
Nevertheless, the mean field solutions contradict the fact that no magnetic
order has ever been observed in the doped copper oxides.  From the
theoretical point of view, it is then very important to study
fluctuation effects in order to determine whether the models concerned are
really relevant to the physics of high temperature superconductors.
Based on a phenomenological
long--wavelength approach for both the spin and charge
degrees of freedom Shraiman and Siggia
\cite{shraiman_tj_spiral2,shraiman_tj_spiral4}
 investigated fluctuations in the
spiral phase. They proposed   short--range spiral order
based on a dynamical mixing of charge and
(out--of--plane) spin fluctuations introduced phenomenologically.

In this paper the issue of quantum fluctuations in the spiral magnetic
phase will be addressed on a microscopic basis, using a functional integral
method.  We provide a systematic investigation of the mechanism which
couples the spin and charge fluctuations in a doped Hubbard antiferromagnet.
The renormalized spin waves are found to be strongly anisotropic in
momentum space. However,
despite of a line of near--zero energies observed in the Goldstone modes, we
show that the spin waves alone are not sufficient to destroy the long--range
order, and the possibility of disordered phases depends crucially on the
incoherent particle--hole excitations.  We shall also demonstrate that the
Hubbard model presents an intrinsic instability towards an inhomogeneous
state (phase separation, CDW, ...) at weak doping,
due to strong coupling between charge
fluctuations and in--plane--spiral fluctuations.  Although phase separation
is suppressed by long--range Coulomb interactions, a possible inhomogeneous
(CDW) state can only be suppressed for sufficiently strong Coulomb
interactions.

We start by writing the action for the Hubbard model in two dimensions
using a local spin reference axis\cite{schulz_path_hubbard}
\begin{eqnarray}
S(\bar{\Phi},\Phi,R ) & = &
\int_0^\beta\!\!d\tau \!\{ \sum_{\bf r} \bar{\Phi}_{\bf r}
( \partial_{\tau} - \mu + \bar{R}_{\bf r} \dot{R}_{\bf r} ) \Phi_{\bf r}
+ H(\bar{\Phi},\Phi,R ) \} \;\;, \label{eq:action2} \\
H(\bar{\Phi},\Phi,R )& =  &
-t \sum_{\langle
{\bf r} {\bf r}' \rangle } (\bar{\Phi}_{\bf r}
\bar{R}_{\bf r} R_{\bf r^\prime} \Phi_{\bf r^\prime} + c.c. )\nonumber\\
& & + \frac{U}{4} \sum_{\bf r} [ (\bar{\Phi}_{\bf r} \Phi_{\bf r})^2
- (\bar{\Phi}_{\bf r} \sigma_z \Phi_{\bf r} )^2 ]  \;\;,
\end{eqnarray}
where $R$ defines the SU$(2)$ rotation from the original electron operators
$\Psi=(\psi_{{\bf r}\uparrow},\psi_{{\bf r}\downarrow})^T $
to the new operators
$\Phi$ via $\Psi=R\Phi$.  $R$ satisfies
$R_{\bf r}(\tau)\sigma_zR^+_{\bf r}(\tau)=
{\bf\Omega}_{\bf r}(\tau)\cdot\bbox{\sigma}$, where
$\bbox{\sigma}=(\sigma_x,\sigma_y,\sigma_z)$ are the Pauli matrices, and
${\bf\Omega}_{\bf r}(\tau)$ is a dynamical variable that defines the local
spin reference axis varying in time and space.

By the saddle point approximation, we can readily identify the spiral phase
as favored among homogeneous candidates. In particular, it is the diagonal
spiral phase with ${\bf Q}=(Q,Q)$
which minimizes the free energy in the strongly
correlated regime ($U\gg t$).  The spiral pitch $p=\cos\frac{Q}{2}$ varies
continuously from one at half--filling to zero in the ferromagnetic limit
above some critical doping.\cite{schulz_path_hubbard} Close to half filling,
we can write $p= 2tx/J $, where $J=4t^2/U$ and $x$ is the hole
concentration.  Within the framework of the Hubbard model, the formation of
the spiral state makes it possible for the charge carriers to propagate
freely, and the quasiparticle structure is determined by a single parameter
$\gamma=\frac{J}{t}(p^{-1}-p)$, which measures the relative strength between
intra-- and inter--sublattice hoppings ($tp$ vs $J(1-p^2)$). As a
consequence, once holes are introduced the Fermi surface immediately shrinks
to a single small pocket around ${\bf k}_m=(\pi/2,\pi/2)$.  The quasiparticle
mass satisfies $ m_-/m_+=1+2\gamma $ within the weakly doped regime, and
thus shows an important anisotropy. Typically, the mass along ($1,1$)
direction $m_+ $ of $\sim 1/J$, much smaller than the mass in its
perpendicular direction $m_-$.

To explore the stability of the spiral phase, we have studied the low energy
(transverse) spin excitations within the one--loop scheme. Assuming
$b_{\bf r}=(\alpha_{\bf r}+i\beta_{\bf r})/2$, with
$\beta_{\bf r}$ ($\alpha_{\bf r}$) describing the
local angular spin deviations within (out of) the spiral spin plane, the
effective spin action becomes
\begin{equation}
S_{eff}
=\frac{1}{4}\sum_{{\bf q}\nu} (\bar{\alpha}_q,\bar{\beta}_q)
\left( \begin{array}{cc}
h_{\bf q} +\Delta_{\bf q}+S_+({\bf q},i\omega_\nu) &
i\omega_\nu +S_z({\bf q},i\omega_\nu) \\
i\omega_\nu +S_z({\bf q},i\omega_\nu) &
h_{\bf q} -\Delta_{\bf q}+S_-({\bf q},i\omega_\nu)
\end{array}\right )
\left( \begin{array}{c}\alpha_q\\ \beta_q \end{array}\right )
\label{eq:Scorr1}
\end{equation}
where
\begin{eqnarray}
h_{\bf q} & = & \sum_{{\bf k}}f_{\bf k}
\{ -g({\bf k},{\bf q})+\frac{2\varepsilon^2_{\bf k}}{U}
 -\frac{g^2({\bf k},{\bf q})}{U}
\},\\
\Delta_{\bf q} & = & 2\sum_{{\bf k}}f_{\bf k}
\frac{\varepsilon_{\bf k}\varepsilon_{{\bf k}+{\bf q}}}{U},
\end{eqnarray}
with $f_{\bf k}$ the Fermi distribution function of the lower Hubbard band,
$\varepsilon_{\bf k}=2t\sin\frac{Q}{2}(\sin k_x +\sin k_y)$ and
$g({\bf k},{\bf q})=2tp(\cos(k_x+q_x)+\cos(k_y+q_y)-\cos k_x-\cos k_y) $.  $
S_\pm({\bf q},i\omega_\nu)$, $S_z({\bf q},i\omega_\nu)$
contain the renormalization of
the spin wave propagator by particle--hole excitations within the lower
Hubbard band.  Without giving detailed expressions,\cite{zhou_unpublished}
we only mention that (i) $S_z({\bf q},i\omega_\nu)$ is an odd function of
$\omega$ and vanishes in the static limit; (ii) more importantly, a
strong anisotropy exists in the interaction vertices
$U_\pm({\bf k},{\bf q})$, which
couple particle--hole excitations across the Fermi surface to the out--of--
(in--)plane spin fluctuations respectively.  In particular, for small doping
the coupling is maximal (minimal) for excitations propagating along
(perpendicular to) the spiral twisting.

The spin wave spectrum is determined by zeros of the determinant of the
$2\times 2$ matrix in eq.(\ref{eq:Scorr1}).  The spectrum contains three
Goldstone zeros at ${\bf q}_0=0,\pm{\bf Q}$, as would be expected from symmetry
considerations.\cite{rastelli_helimagnet}  In agreement with previous
findings,\cite{gan_tj_spiral} to correctly account for the spin wave
structure in the spiral state, it is crucial to take into account the
dynamical effects of particle--hole excitations
($h_{\bf q}+\Delta_{\bf q}=-S_+({\bf q},0)\ne 0$ at ${\bf q}_0=\pm{\bf Q}$).

Fig.\ref{fig:mode} shows the numerical results for the spin wave spectrum
within the Brillouin zone.  In comparison with the half--filled case where
particle--hole excitations are absent in the lower Hubbard band, the overall
spectral structure has changed drastically for even the slightest doping.
Summarizing the numerical results, we find: (i) spin waves are observed
in the ${\bf q}\rightarrow 0$ limit only for very weak doping, they are
completely dissolved into the particle--hole excitation spectrum as doping
further increases.  On the other hand, the spin wave velocity is always
slightly above the half filling value of $\sqrt{2}J$ when the mode is not
overdamped, which disagrees with the result derived from t--J
model\cite{gan_tj_spiral} but agrees qualitatively with Chubukov and
Frenkel;\cite{chubukov_rpa} (ii) around $\pm{\bf Q}$, spin wave excitations
are
strongly renormalized.  In particular, they have very low energies along the
diagonal direction for wave vectors between $\pm({\bf Q},{\bf Q}_0)$,
a region of length $\sim 4Ux/t$.
These excitations will certainly dominate the low
energy properties in the spiral state; (iii) more surprisingly, an extra
zero is found in the spectrum immediately after the spin waves reemerge from
the the particle--hole pair spectrum (at ${\bf q}^\prime\ne 0$).  This
unexpected
result will be seen to signal the instability of the spiral state against an
inhomogeneous state.

Things are rather transparent in the long wavelength limit.  Increasing
doping raises the Fermi velocity,
which in turn pushes up the upper bound of the pair spectrum.
The spin wave excitations near ${\bf q}=0$ then become overdamped as doping
exceeds a certain threshold. Below the threshold the leading correction $
S_-({\bf q}\!\rightarrow\!0,\omega\!\rightarrow\!0)|_{\omega/|{\bf q}|=c}$ has
its
origin in the coupling between spin fluctuations {\it within} the spiral
plane and particle--hole excitations just across the Fermi surface, which
scales as $x^{1/4}$ as doping approaches zero.  Correspondingly, the spin
wave velocity has approximately the form:
\begin{equation}
c=\sqrt{2}J + J\sqrt{\frac{2\pi t}{J}}
\left(\frac{q_x+q_y}{\sqrt{q^2_x+q^2_y}}\right)^3x^{\frac{1}{4}}
\label{eq:vitesse0}
\end{equation}
This disagrees with Gan {\it et al.} \cite{gan_tj_spiral} who find a leading
correction of $O(x)$.  From eq.(\ref{eq:vitesse0}), it is clear that
renormalization in the spin wave velocity is maximal along $(1,1)$
direction, and minimal in its perpendicular direction.

For the dynamical spin susceptibility near $\pm{\bf Q}$, out--of--plane spin
fluctuations dominate. Here, mixing with charge excitations is crucial in
determining the low--lying spin excitations.  We have already indicated that
the Goldstone modes at $\pm{\bf Q}$ appear as the direct consequence of the
spin--charge coupling. Around $\pm{\bf Q}$, $S_+({\bf q},\omega)$ dominates
the selfenergy correction, and the spin wave spectrum can be approximated
by:
\begin {equation}
\omega^2_{\bf q}/J^2=(1+\frac{4}{\gamma}x)(\delta q_x-\delta q_y)^2+
4x(1+\frac{1}{\gamma})(\delta q_x+\delta q_y)^2
\label{eq:specQ}
\end{equation}
provided that the hole concentration is small but finite (such that $tp$ or
$\gamma$ remains finite).  Eq.(\ref{eq:specQ}) corresponds to the so--called
torsion mode in Shraiman and Siggia's
description.\cite{shraiman_tj_spiral2,shraiman_tj_spiral4}.  We emphasize
the strong anisotropy observed in the spin wave spectrum (see also
Fig.\ref{fig:mode}b).  In particular, the characteristic velocity is
$O(\sqrt{x}J)$ in the $(1,1)$ direction where the renormalization is maximal
and $O(J)$ otherwise.

Finally, the unexpected extra zero in the spin wave spectrum (at
${\bf q}^\prime$) belongs, in fact, to an imaginary mode existing for
${\bf q}\in(0,{\bf q}^\prime)$,\cite{zhou_unpublished} which certainly implies
the
instability of the spiral phase.  However, what is remarkable here is its
{\em simultaneous} presence with the Goldstone modes.

That in--plane and out--of--plane spin fluctuations are decoupled in the
static limit helps in determining both the nature of this instability and
the offending fluctuations.  In fig.\ref{fig:SmSp} we have plotted the
variation of $S_-({\bf q},0)$ for a given $U/t$ and a hole density.  The
presence
of a sharp peak in $S_-({\bf q},0)$ easily explains the qualitative change in
the
nature of spin fluctuations {\it within} the spiral plane at sufficiently
low doping.  In this regime, since $h_{\bf q}-\Delta_{\bf q}$ is a smooth
function and
approaches zero in the limit ${\bf q}\rightarrow 0$, we readily find that under
the condition:
\begin{equation}
U\rho(0) >\frac{q^2_x+q^2_y}{(q_x+q_y)^2}
\label{eq:criterion}
\end{equation}
the spin propagator $\langle\beta_q\bar{\beta}_q\rangle_{\omega\rightarrow
0}$ changes sign for wavevectors ${\bf q}\in(0,{\bf q}^\prime)$, and
correspondingly
the spin stiffness becomes negative for fluctuations inside the spiral
plane.  Here $\rho(0)$ is the quasiparticle density of states at the Fermi
energy,

The nature of the instability at ${\bf q}=0,{\bf q}^\prime$ is further
clarified by
studying the charge response to the corresponding spin fluctuations.  A RPA
calculation for the density--density response function $\chi({\bf q},\omega)$
reveals that the static charge susceptibility couples only to the spin
fluctuations within the spiral plane.  The divergence in
$\langle\beta_q\bar{\beta}_q\rangle_{\omega\rightarrow 0}$ then leads to a
similar behavior in the charge susceptibility. As a result we find in the
limit ${\bf q}\rightarrow 0$ a negative compressibility under exactly the same
condition, eq.(\ref{eq:criterion}), which predicts phase separation. This has
been realized in earlier studies of the compressibility based on mean field
solutions.\cite{auerbach_inhmphase,arrigoni_incomm} Further, the divergence
of $\chi({\bf q},0)$ at a finite wavevector clearly indicates a CDW
instability.
Contrary to phase separation which is often overemphasized,
we note that this second instability has received little attention
in previous studies.
On the other hand, for the wide parameter
region we have examined, we have not seen any indication of a non--coplanar
phase claimed by Chubukov and Musaelian.\cite{chubukov_unpublished}

Of course, phase separation would hardly survive in real materials in view
of long--range interactions neglected in the Hubbard model.  We have
therefore reexamined the problem in the presence of a $1/r$ potential. We
find that, in the static limit, Coulomb interactions modifies {\it only} the
spin fluctuations {\it within} the spiral plane, (and along with them, the
charge density response function). The screened spin propagator
$\langle\beta_q\beta_{-q}\rangle_{\omega\rightarrow 0}$ now becomes:
\begin{equation}
\langle\beta_q\beta_{-q}\rangle_{\omega\rightarrow 0}\approx
\frac{-2}{h_{\bf q}-\Delta_{\bf q}+\frac{1}{2}U^2_{f-}\chi_0({\bf
q},0)/\epsilon({\bf q},0)}
\label{eq:betaR}
\end{equation}
where $\chi_0({\bf q},0)$ is the bare charge susceptibility in the lower
Hubbard
band, and $\epsilon({\bf q},0)=1-V_{\bf q}\chi_0({\bf q},0)$ is the dielectric
constant in
the presence of the Coulomb potential $V_{\bf q}$.  Eq.(\ref{eq:betaR}) has
been
obtained by assuming that the interaction vertex $U_-({\bf k},{\bf q})$ can be
approximated by its value on the Fermi surface, $U_{f-}({\bf q})$, which is
correct at sufficiently weak doping.  From eq.(\ref{eq:betaR}) we easily see
that even weak Coulomb interactions stabilize against phase separation.  On
the other hand, the instability against a CDW state can only be eliminated
for sufficiently strong Coulomb interactions.
In the intermediate regime, a possible compromise can be established
by forming a modulated spiral phase, or domain walls, as suggested by
Dombre.\cite{dombre_domainwall}

A physically interesting possibility is the disordering of the long--range
ordered spiral state induced by low--lying (torsion) modes, as suggested by
Shraiman and Siggia.\cite{shraiman_tj_spiral2,shraiman_tj_spiral4} In what
follows, we assume the instabilities discussed above to be absent in real
systems, and for that purpose suppress by hand the corresponding vertex
$U_-({\bf k},{\bf q})$.\cite{footnote1} Then one can verify that the imaginary
mode
and hence the extra zero in the spin wave spectrum disappear. Otherwise, the
low--lying Goldstone modes remain nearly unchanged.\cite{zhou_unpublished}
We have calculated the renormalized spin amplitude (per electron)
$m=1/2-|b|^2 $ in the presence of these zero point fluctuations.  In
general, $|b|^2\equiv(|\alpha|^2+|\beta|^2)/4$ consists of contributions
from the spin wave modes as well as the incoherent particle--hole
excitations due to their coupling with spin fluctuations. By analogy with
the half--filled case we concentrate on the spin--wave corrections.
Fig.\ref{fig:sz} shows $m$ as a function of doping for two values of
$U/t$. The variation of $m$ reflects the magnetic frustration of the system
upon doping. Starting from the N\'eel state at half filling, where
$m=S-0.197$, doping initially induces frustration leading to the spiral
phase and to an increase in zero--point spin fluctuations. However, after
reaching a maximum (where $m$ is minimized), frustration reduces
continuously until the ferromagnetic limit, where no zero--point
fluctuations are expected and $m$ remains unchanged at its saturation value.
The $U/t$ dependence of $m$ can also be understood.  Provided that the
maximal frustration is reached at a certain spiral pitch $p_c$, the optimal
doping $x_c$ where $m$ is minimized decreases as $U/t$ increases. In
particular, $x_c$ varies inversely proportional to $U/t$ in the weak doping
regime. From the above calculation we have learned that even in the most
frustrated case where zero--point spin fluctuations are the most violent
they are not sufficient to destabilize the spiral phase in favor of a
disordered state.

To conclude, our analysis on the quantum fluctuations in a spiral phase
reveals an intrinsic instability in the weakly doped Hubbard model.
Phase separation is suppressed by
the long--range Coulomb interactions in a real
material, however, we emphasize that an inhomogeneous state may well
persist
provided the long--range interactions were not too strong. A
possible
scenario is then the formation of a structure similar to the domain walls
found in the weak correlation regime,\cite{dombre_domainwall}
before the system eventually conducts.
We remark that such a scenario may be relevant the recent observations
in
La$_{\text 2}$NiO$_{\text{4+y}}$\cite{tranquada_LaNiO}.
Within the present framework, we have
further examined the possibility of disordered spiral phases. We have shown
that zero--point spin wave excitations alone are not sufficient to destroy
the long--range magnetic order. Nevertheless, the contribution from the
incoherent excitations in the system grows as one further dopes the
system. We expect that above a critical doping, an inhomogeneous state
should finally give way to a disordered state with strong incommensurate
correlations.  Finally, the hole dynamics in such doped systems is far from
being understood.

\begin{figure}\caption{(a) Spin wave spectrum in the $(1,1)$ spiral state.
for $U/t=10$ and $x=1.5\%$(above), $x=7.5\%$(below). The region
between doted lines stands for the continuum for particle--hole excitations.
(b) The spectrum near
$\pm{\bf Q}$ for $U/t=20$ and $2\%$ doping.}  \label{fig:mode}
\end{figure}
\begin{figure}\caption{ The selfenergy correction for the static spin
propagator within the spiral plane, $S_-({\bf q},0)$, for $U/t=20$ and
$x=5\%$.}
\label{fig:SmSp}
\end{figure}
\begin{figure}\caption{Renormalized spin amplitude $m$
as a function of doping in the presence of
spin wave excitations.}
\label{fig:sz}
\end{figure}


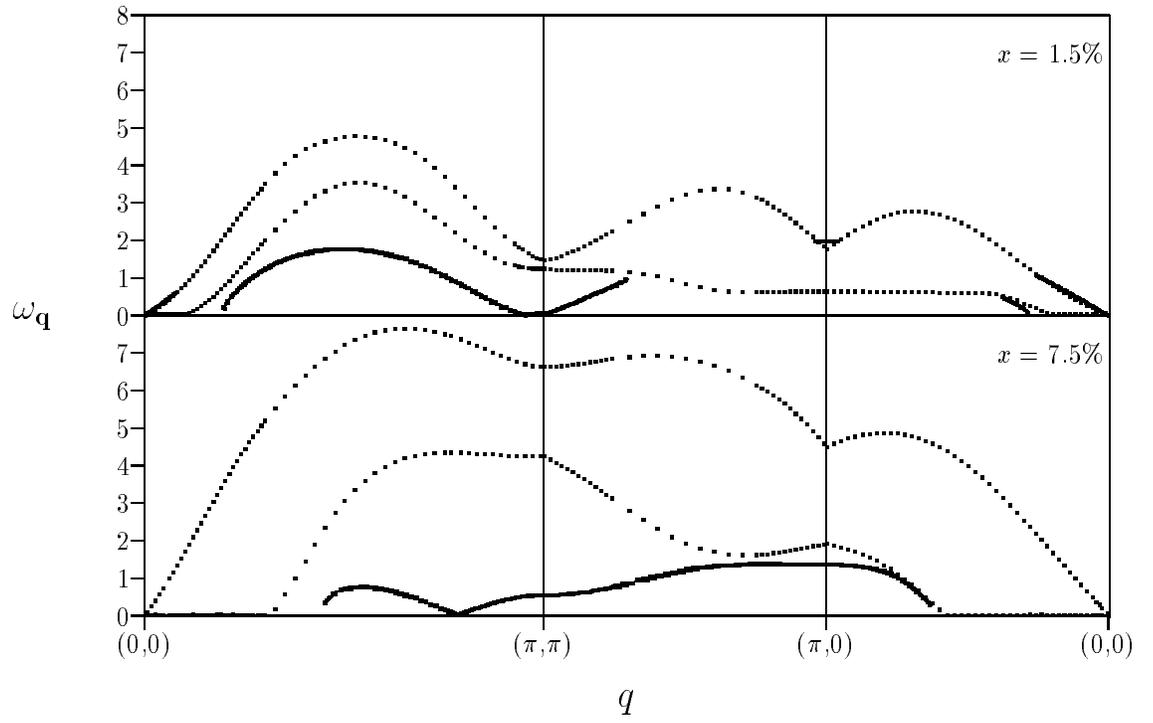
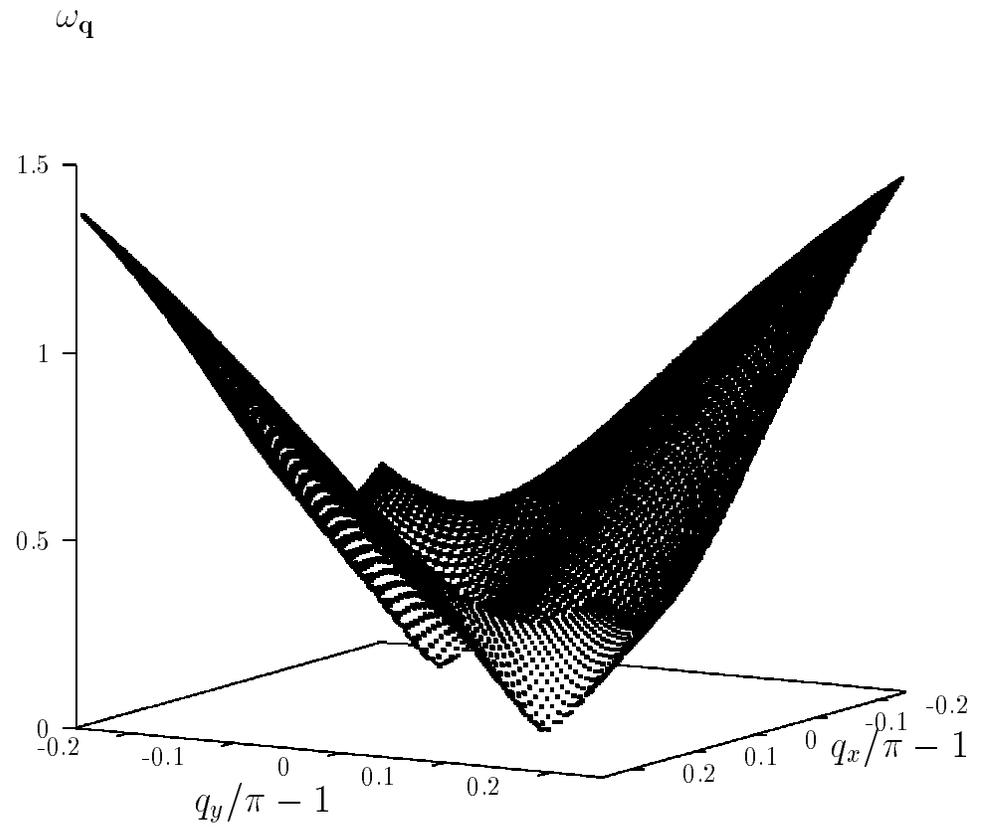
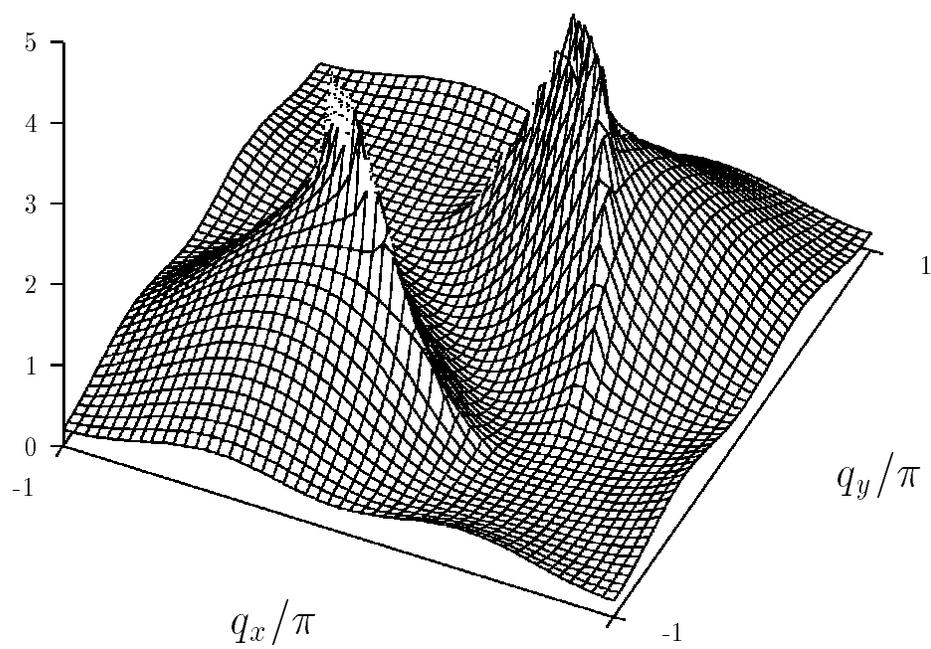
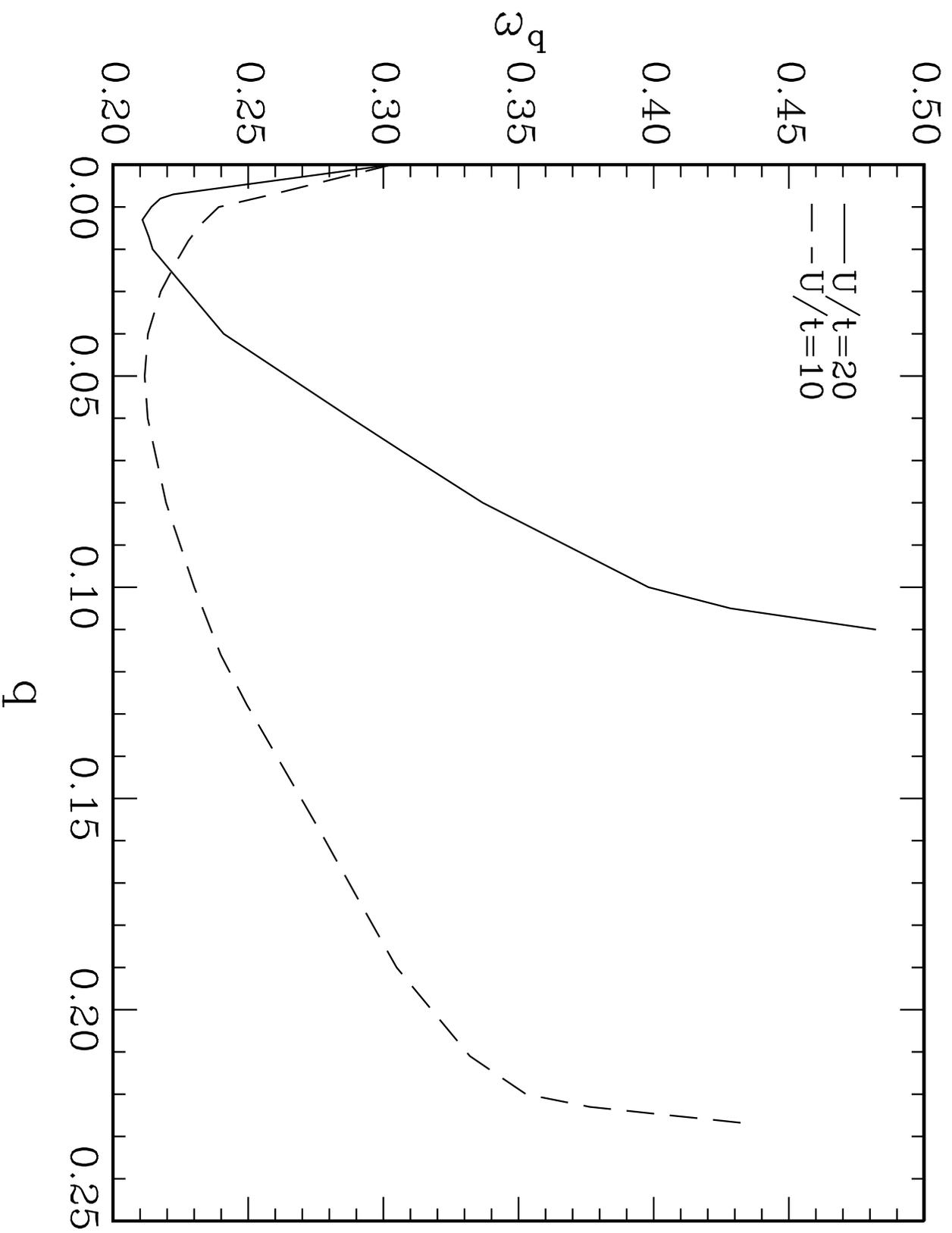
\end{document}